\documentclass[aps,prd,twocolumn,showpacs,superscriptaddress]{revtex4}

\usepackage{times} 
\usepackage{graphicx}
\usepackage{amsmath,amsthm,amssymb}
\usepackage{amssymb}
\usepackage{bm}
\usepackage{hyperref}
\usepackage{float}
\usepackage[usenames, dvipsnames]{color}

\def\apj{Astrophysical Journal}
\def\mnras{Monthly Notes of Royal Astronomical Society}

\def\aj{Astronomical Journal}
 
\def\aap{Astronomy and Astrophysics}     
\def\prd{Phyical Review D}
\def\prl{Phyics Review Letter}
\def\plb{Physic Letter B}
\def\pasa{Publications of the Astronomical Society of Australia}
\def\jhep{Journal of High Energy Physics}
\def\jcap{Journal of Cosmology and Astroparticle Physics}

\begin{document} 

\title{Ghostly Galaxies as Solitons of Bose-Einstein Dark Matter }

\author{Tom Broadhurst}
\email{tom.j.broadhurst@gmail.com}
\affiliation{Department of Theoretical Physics, University of the Basque Country UPV/EHU,
             E-48080 Bilbao, Spain}
\affiliation{Donostia International Physics Center (DIPC), 20018 Donostia-San Sebastian (Gipuzkoa)
	Spain.}
\affiliation{Ikerbasque, Basque Foundation for Science, E-48011 Bilbao, Spain}

\author{Ivan De Martino }
\email{ivan.demartino@dipc.org}
\affiliation{Donostia International Physics Center (DIPC), 20018 Donostia-San Sebastian (Gipuzkoa)
	Spain.}

\author{Hoang Nhan Luu}
\email{hnhanxiii@gmail.com}
\affiliation{Institute for Advanced Study and Department of Physics, Hong Kong University of Science and Technology, Hong Kong}

\author{George F. Smoot}
\email{}
\affiliation{Institute for Advanced Study and Department of Physics, Hong Kong University of Science and Technology, Hong Kong}
\affiliation{WF Chao Foundation Professor, IAS, Hong Kong University of Science and Technology, Clear Water Bay, Kowloon, 999077 Hong Kong}
\affiliation{Paris Centre for Cosmological Physics, APC, AstroParticule et Cosmologie, Universit\'{e} Paris Diderot, CNRS/IN2P3, CEA/lrfu}
\affiliation{Universit\'{e} Sorbonne Paris Cit\'{e}, 10, rue Alice Domon et Leonie Duquet, 75205 Paris CEDEX 13, France}

\author{S.-H. Henry Tye }
\email{iastye@ust.hk}
\affiliation{Institute for Advanced Study and Department of Physics, Hong Kong University of Science and Technology, Hong Kong}
\affiliation{Department of Physics, Cornell University, Ithaca, NY 14853, USA}

\date{}

\begin{abstract}
 The large dark cores of common dwarf galaxies are unexplained by the standard heavy particle interpretation of dark matter. 
 This puzzle is exacerbated by the discovery of a very large but barely visible, dark matter dominated galaxy Antlia II orbiting the Milky Way, uncovered by tracking star motions with the {\t Gaia} satellite. Although Antlia II has a low mass, its visible radius is more than double any known dwarf galaxy, with an unprecedentedly low density core. We show that Antlia II favors dark matter as a Bose-Einstein condensate, for which the ground state is a stable soliton with a core radius given by the de Broglie  wavelength. The lower the galaxy mass, the larger the de Broglie wavelength, so the least massive galaxies should have the widest soliton cores of lowest density. An ultra-light boson of $m_\psi \sim 1.1  \times10^{-22}$ eV, accounts well for the large size and slowly moving stars within Antlia II, and agrees with boson mass estimates derived from the denser cores of more massive dwarf galaxies. For this very light boson, Antlia II is close to the lower limiting Jeans scale for galaxy formation permitted by the Uncertainty Principle, so other examples are expected but none significantly larger in size. This simple explanation for the puzzling dark cores of dwarf galaxies implies dark matter as an ultra-light boson, such as an axion generic in String Theory.
\end{abstract}

\maketitle

Dark matter (DM) is understood to be non-relativistic even in the early Universe, otherwise initial density perturbations destined to become galaxies would be smoothed away by free streaming of the dark matter. This cold dark matter (CDM) has long been synonymous with heavy particle interpretations beyond standard particle physics, but no such particles have been detected in stringent laboratory experiments\cite{Xenon1T}. Black holes also qualify as CDM, and although LIGO/Virgo has claimed an abundance of $30 M_\odot$ black holes, their space density is limited to less than 5\% of all dark matter by micro-lensing measurements through massive lensing clusters  \cite{Kelly}.  { Wang et al. (2018) \cite{Wang2018} used LIGO detection to obtain a tight upper limit on the abundances of primordial black holes demonstrating that they can only contribute to a level of a few percent to the dark matter abundance, which has been independently confirmed in \cite{2018ApJ...857...25D,2018PhRvD..97b3518O}.  Finally, and more in general, bounds placed on the abundance of MACHO content in the Galactic halo from microlensing surveys such as MACHO, OGLE, and EROS, rule out any their significant contribution \cite{machos2007}. }

Enthusiasm for CDM has long been tempered by the shallow mass profiles of common dwarf galaxies that appear to be cored rather than ``cuspy" as predicted by N-body simulations. 
It has been hard to make sense of this apparent contradiction in the context of CDM without invoking hypothetical forces or implausibly transformational gas outflows. The very low density of stars in Antlia II and their low metallicities argues empirically against an explanation that repeated episodes of star formation somehow flatten a CDM cusp into a core \cite{Pontzen2012,Read2016}, an idea  that is { disfavoured} by accurate high resolution simulations \cite{Bose2018}. 

The above contradiction does not arise for a very different form of non-relativistic dark matter, as a Bose-Einstein condensate. Pioneering simulations in this context reveal dark cores of very light bosons corresponding to the ground state of this quantum, wave-like form of dark matter\cite{Schive2014a}, termed $\psi $DM. These $\psi $DM simulations simply evolve a coupled Schr\"odinger-Poisson equation describing a non-relativistic self-gravitating condensate \cite{Widrow1993, Hu2000} requiring only one free parameter, the boson mass, $m_\psi$, where the smaller $m_\psi$ the larger the de Broglie wavelength. Rich non-linear structure is revealed by the $\psi $DM simulations on the de Broglie scale, including a prominent, standing wave core at the center of every galaxy that is a stable soliton, representing the ground state, surrounded by a halo of turbulent, self interfering excited states \cite{Schive2014a,Schive2014b} and confirmed by independent simulations\cite{Veltmaat2018}.

Here we compare Antlia II with the central prediction of $\psi$DM, that the least massive galaxies should have the widest soliton cores of lowest density. The de Broglie wavelength is of course larger at lower momentum, so the soliton radius depends inversely with soliton mass, $R_{sol} \propto M_{sol}^{-1}$. Also the $\psi$DM simulations have established $M_{sol}$ increases with the total galaxy mass, as $M_{sol} \propto M_{gal}^{1/3}$ \cite{Schive2014b, Veltmaat2018} so the central DM density of soliton cores scales with galaxy mass as:

\begin{equation}
    \overline{\rho}_{sol} =  1.9\times10^{6}\biggl(\frac{M_{gal}}{10^9 M\odot}\biggr)^{4/3} \biggl(\frac{m_\psi}{10^{-22} ~ \rm{eV}}\biggr)^2 \quad \frac{M_\odot}{\rm{kpc}^{3}}
\end{equation}

Hence, at fixed $m_\psi$ the core density of a dwarf galaxy of $10^9 M_\odot$, should be much smaller, $10^{-4}$, than the central DM density of a massive galaxy of $10^{12} M_\odot$, like the Milky Way. This contrasts with a predicted {\t increase} in density of $\simeq 30$ for standard CDM, from the ``concentration-mass relation" of N-body simulations, where lower mass galaxies are predicted to have denser dark matter profiles. 

Despite the general tendency to accommodate CDM, all well studied dwarf spheroidal (dSph) galaxies are consistently claimed to have large dark cores, traced by a diffuse distribution of old stars of low velocity dispersion. In particular, the best studied Fornax dSph is determined by several different methods to have a core radius of $\simeq 1.0 kpc$ \cite{Amorisco,Binney2018} with a density profile that is accurately fitted by the soliton form \cite{Schive2014a} of $\psi $DM, as shown in Figure 1, for which a Jeans analysis yields a boson mass of, $m_\psi=0.8\pm0.2\times 10^{-22}$ eV \cite{Schive2014a}. Furthermore, Fornax provides another independent and compelling argument for $\psi $DM implied by the presence of ancient globular clusters on large orbits around Fornax. This is unexpected for discrete dark matter, such as CDM or black holes, that would be focused gravitationally by an orbiting globular cluster into a "wake", so the globular clusters should have migrated long ago to the center of Fornax \cite{Goerdt2006,Cole2012}. This ``dynamical friction" is not significant for $\psi$DM, which cannot be confined to less than the de Broglie scale because of the Uncertainty Principle \cite{Lora2012, Hui2016}, leaving the Fornax globular cluster orbits little affected \cite{Hui2016}. 
\begin{figure}[!ht]
\centering	
\includegraphics[width=0.99\columnwidth]{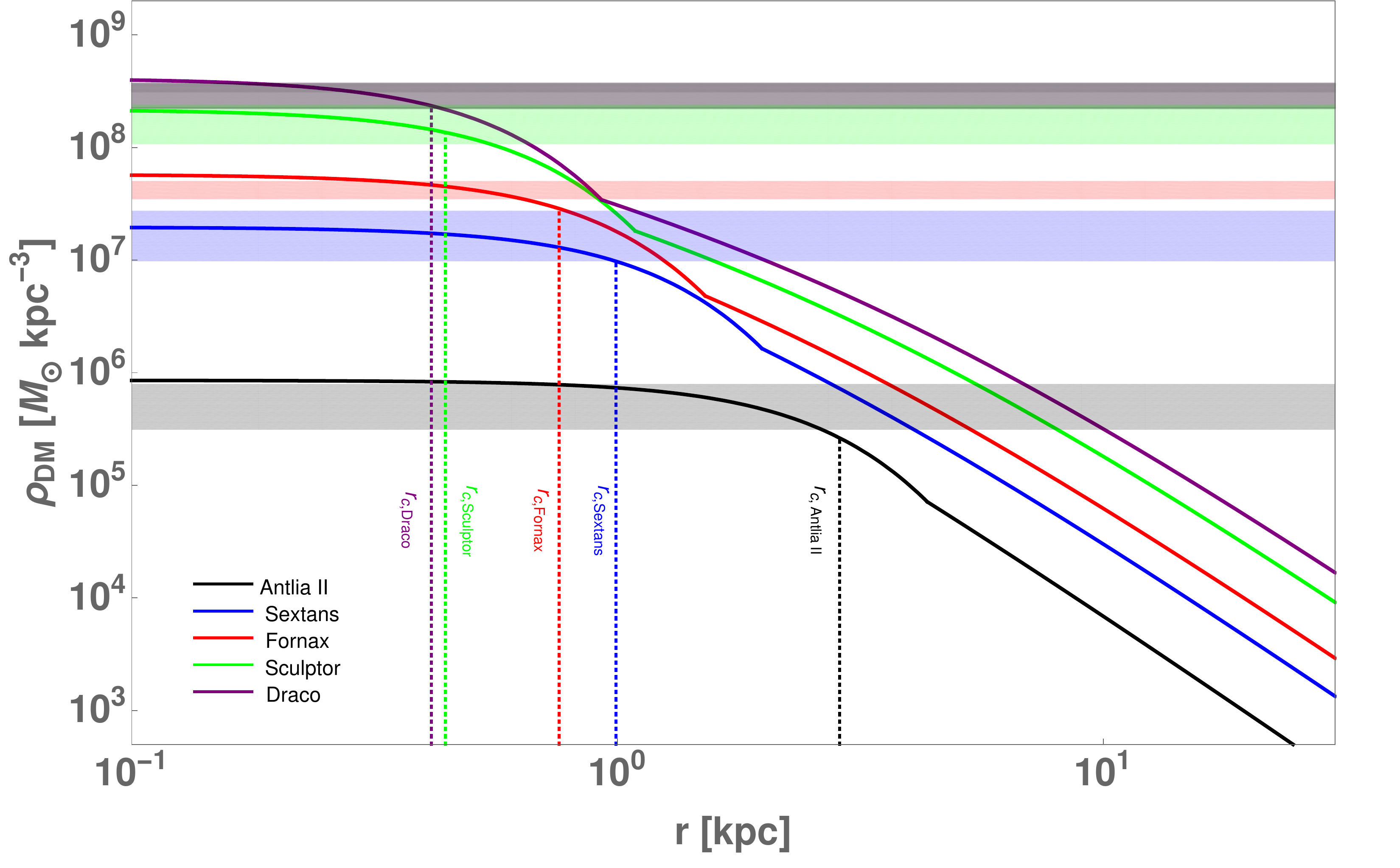}
\caption{Here we show how the low mean density reported for Antlia II is readily reproduced by the wide solitonic core density profile of a low mass galaxy of $\sim 10^9M_{\odot}$ in the context of $\psi$DM. For comparison we compare the reported core mass densities of other well studied dwarf spheroidal galaxies with the density profiles for $\psi$DM, for the same boson mass, $m_\psi=10^{-22}$ eV, demonstrating the consistency with the family of predicted $\psi$DM profiles, where soliton radius and mass are inversely related.}\label{fig1}
\end{figure}

 The newly discovered Antlia II galaxy has an exceptionally large core radius of $\simeq 3$ kpc, which together with its small velocity dispersion $\simeq 6$ km/s, corresponds to a mean DM density of only $\simeq 10^5M_\odot {\rm kpc}^{-3}$, that is an order of magnitude lower than any known dwarf galaxy \cite{Torrealba2018} and $\simeq 30$ times lower than Fornax, as shown in Figure~1. Antlia II extends the trend of recent discoveries towards larger, lower surface brightness galaxies of low velocity dispersion, including Crater II and other large dSph galaxies in orbit around Andromeda \cite{Collins2013,Torrealba2016,Caldwell2017}. At face value these "ghostly" galaxies are encouraging for the $\psi $DM interpretation of DM, particularly Antlia II\cite{Torrealba2018}, and so here examine 
 whether the well studied dwarf spheroidals follow the distinctive density-radius
 relation of eqn~1, that has the {\it opposite} sign to the behaviour expected for CDM. { It will also be of interest to examine the inner regions of more massive galaxies, requiring high resolution to resolve the smaller, soliton scale predicted of $\simeq 100$ pc. It is encouraging that careful treatment of complex "3D" gas dynamics of intermediate mass galaxies reveals evidence of steep inner profiles and much smaller cores than typically derived using standard disk models \cite{Kurapati2020}.}
 In Figure~1 we show 
 a family of $\psi$DM profiles as a function of galaxy mass, following eqn~1, with the boson mass $m_\psi$ set to the canonical $10^{-22}$ eV, appropriate for $\psi$DM \cite {Hu2000,Schive2014a,Chen2017}. These model profiles can be seen to match the reported mean densities of the well studied dwarf spheroidal, listed in Table 1 (Appendix \ref{app:C}) showing that Antlia II { is fully consistent with a solitonic core} that is 2.5 times larger than Fornax mass and hence 40\% lower mass than the core mass of Fornax, of $5\times10^7 M_\odot$, and in good agreement with the mass estimated by Torrealba et al. (2018) based on the observed velocity dispersion of $6.5$ km/s at the half light radius of 2.8 kpc.
\begin{figure}[!ht]
	\centering
\includegraphics[width=0.98\columnwidth]{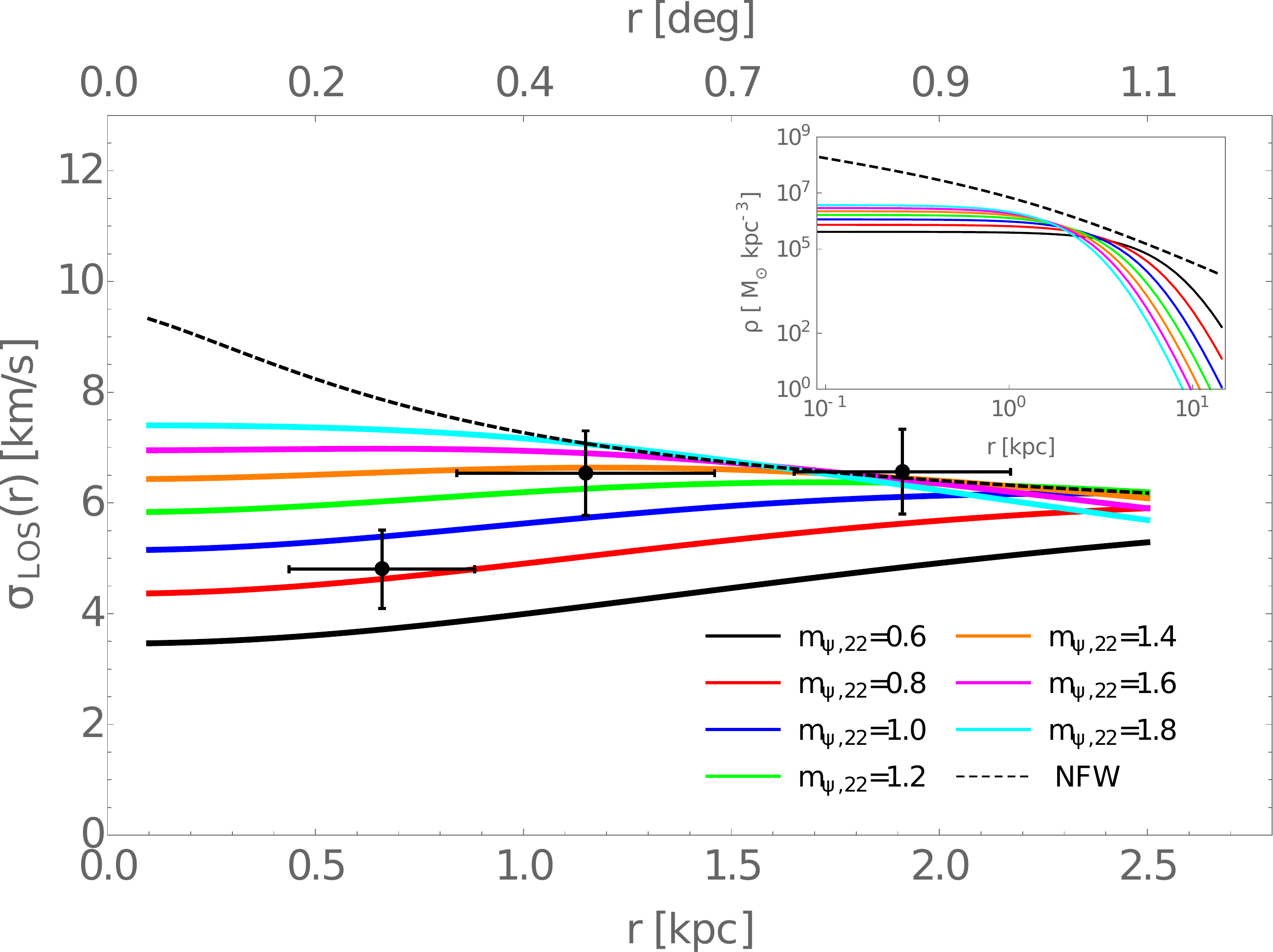}
	\caption{The measured velocity dispersion (data points) \cite{Torrealba2018}, compared with the predictions of $\psi$DM mass profiles for a range of light boson masses - see legend where $m_{\psi,22}\equiv m_\psi/10^{-22}$. The central decline in dispersion velocity for the lighter bosons matches well the data, and reflects the low central mass density. This contrasts with the relatively more concentrated best fitting NFW profile (dashed curve) where an enhancement is expected, unlike the data. }\label{fig2}
\end{figure}
  We can also compare the measured radial velocity dispersion profile observed for Antlia II \cite{Torrealba2018} with $\psi $DM  predictions by solving the Jeans equation in projection (see Eq. \eqref{eq:sol_projected}). We assume the commonly adopted Plummer profile appropriate for the stellar distribution of dwarf spheroidal galaxies, normalized to the measured half light radius of 2.8 kpc \cite{Torrealba2018}. The predicted profiles shown in Figure~2 have a slow centrally declining velocity dispersion due to the declining interior mass, $M(<r)$ for constant density cores, with a somewhat steeper decline for low boson mass. A boson mass is favoured of $0.4 - 1.2\times10^{-22}$ eV, (Figure 2 and Appendix \ref{app:C}) below which the observed mean dispersion of $\simeq 6.5$ km/s is under predicted and above which the soliton radius falls short of $2.5$ kpc radius of the outer bin, so the central velocity dispersion exceeds the measured value, shown in Figure 2. The opposite behavior is expected for CDM (Figure 2), where the continuously rising central density predicts a rising velocity dispersion that exceeds the data in Figure 2, for the best fitting NFW profile derived in the analysis of Torrealba et al. (2018) \cite{Torrealba2018}.
  
 We now jointly constrain the boson mass and soliton core radius with galaxy mass in Figure~3. We compare Antlia II with Fornax and Sextans dSph's because no tidal effects have been detected in deep imaging and careful dynamical work\cite{Pena2008,fornaxnotides,sextans_waveanalysis}, so we need not be overly concerned that their masses are underestimated. In any case, these galaxies are established to have large orbits about the Milky Way extending to $\geq 100$ kpc, including Antila II which is presently at a radius of 130 kpc with a sizeable estimated pericenter of 50 kpc implying tidal effects are marginal\cite{subaru,Torrealba2018}. 
 
 \begin{figure}[!ht]
\centering
\includegraphics[width=0.98\columnwidth,height=7.5cm]{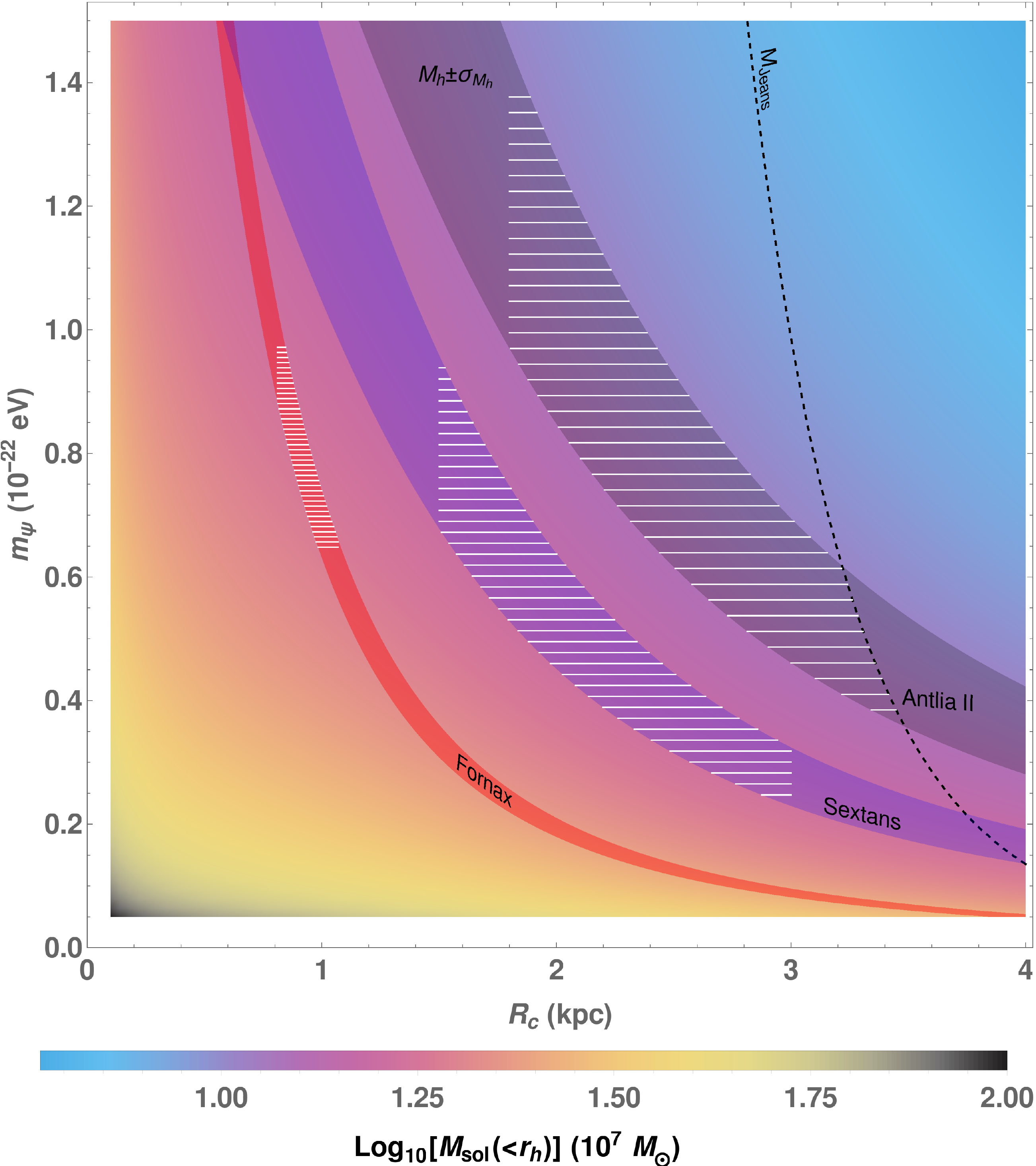}
\caption{The plane of core radius versus $m_\psi$ as a function of galaxy mass (colour coded) predicted for $\psi$DM that are limited to the hatched regions derived from the observations of Antlia II summarized in  Table  1  (Appendix \ref{app:C}). Specifically, we have used the observed core radius to obtain those regions. The Fornax, Sextans dwarf galaxies are included for comparison because their density profiles are understood not to have been modified significantly by tidal stripping.  In this case, the hatched regions are derived using the constraints on the core radius and the boson mass listed in Table  1.  The limiting Jeans mass is also indicated as a dashed black curve that places an upper limit on the radius of Antlia II at fixed $m_\psi$. Despite the wide range of mass and radius these three galaxies which span well over an order of magnitude in density are compatible with the same boson mass, $m_\psi \sim  10^{-22}$ eV.}\label{fig3}
\end{figure}

 In Figure 3 we define a contour corresponding to the measured mass of $5.4\pm 2.1\times 10^7M_{\odot}$ interior to a limiting radius of $\simeq 2.5$ kpc within which the velocity dispersion profile of Antlia II is observed to be flat\cite{Torrealba2018}, as shown in Figure 2, so the soliton core extends to at least this radius. Importantly, Figure 3 shows that despite the considerable differences in mass and core radius between these three galaxies a common boson mass can evidently be inferred, of $\simeq 1.1 \times10^{-22}$ eV within the uncertainties and for
 all dwarf spheroidal galaxies listed in Table 1 that have been analysed in this context (see Appendix \ref{app:C}).

The Uncertainty Principle not only sets the soliton scale above, but also provides another fundamental prediction of a sharp minimum halo mass by imposing a ``quantum Jeans condition" \cite{Sikivie, chavanis2018,Marsh2014} because the DM cannot be confined within the de Broglie wavelength thereby preventing galaxy formation below a limiting Jeans mass (for more details see \cite{Hui2016} and reference therein): 
\begin{align}
  M_{J} = &\frac{3}{2}\Biggl(\frac{\Omega_a h^2_{70}(1+z)^3}{0.27}\Biggr)^{\frac{1}{4}}\Biggl(\frac{m_\psi}{10^{-22}  ~ \rm{eV}}\Biggr)^{-\frac{3}{2}} \,\,\, 10^7 M_\odot\,,  
\end{align} 

as shown in Figure~3, where $h_{70}\equiv (H_0 /70 \rm{\, km \,s}^{-1} \rm{Mpc}^{-1})$ is  the dimensionless Hubble  constant  and  $\Omega_\psi$ is the cosmological fraction of  the  dark matter  critical  density. This limit depends only weakly on formation redshift as the power spectrum cuts off below a Jeans-like wavelength, $\lambda_J \propto (1+z)^{1/4}$ \cite{Marsh2014, Schive2014b,Lee2016}, defining a relatively sharp lower limiting galaxy mass for $\psi $DM \cite{Hu2000,Bozek2015,Schive2014b,chavanis2018}. Hence, low mass galaxies should be abundant towards this limit and firmly absent below it. Furthermore, the presence of this mass limit predicts a maximum soliton core radius of $\simeq 3 $ kpc, as shown in Figure 3, because the lowest mass galaxies should have the widest solitons of lowest mass density, with Antlia II lying closest to this existential limit.

Another ghostly galaxy is the "feeble giant" Crater II, orbiting the Milky Way at $50$ kpc, with a sizeable radius of $1.1 $ kpc and surprisingly low velocity dispersion of only $\simeq 3$ km/s\cite{Torrealba2016}. Most of the DM of Crater II has likely been stripped off tidally as indicated by surface brightness distortions and because of its small inferred pericenter of only 10 kpc \cite{Torrealba2016} where stripping is generally expected to significantly reduce the velocity dispersion and truncate the stellar radius, depending on the phase of the orbit \cite{Pena2008,Sanders2018,FuTidal} and so previously Crater II may have resembled more Antlia II with a larger radius and higher velocity dispersion. Another large dwarf spheroidal galaxy, Cetus  at 800 kpc and is determined to be one of only a few ``isolated" galaxies that has not suffered significant interaction with other local group galaxies \cite{cetus}. This galaxy extends to at least 3 kpc in deep imaging, with no detectable tidal truncation radius \cite{cetus}. The velocity dispersion of Cetus is close to $10$ km/s and can traced beyond its its half light radius to$\simeq 1.4$ kpc\cite{cetus}. The extended stellar profile may indicate a soliton core similar to Fornax (in Figure 1) surrounded by a lower density DM halo as predicted for $\psi$DM in the absence of tidal truncation, extending to several kpc, as shown in Figure 1.

The above large core "classical" dSph galaxies may be contrasted with the newly discovered class of much smaller $20-50$ pc "ultra faint dwarfs" uncovered in wide field surveys\cite{WS,Calabrese2016} on relatively small orbits, $< 50$ kpc, within the Milky Way. These relatively small objects are very DM dominated, given their typical velocity dispersion of $\simeq 3$ km/s and low luminosities, of typically only $\simeq 1000$ L$_\odot$ and so together with their small orbits they are generally considered to be heavily stripped \textquotedblleft remnants" \cite{Mutlu-Pakdil,Longeard,Nadler2018} of originally large dSph galaxies.  In the context of $\psi$DM, tidal stripping is estimated to be significantly more efficient than for CDM, as the soliton core expands in response to the loss of outer stripped halo mass pushing more DM beyond the tidal radius as the soliton expands in radius in response to the reduced mass, in a runaway process\cite{Du2018}. This "remnant" origin for the UDF galaxies may be supported by the serendipitous discovery of central star clusters within several dSph galaxies, with sizes and luminosities similar to the UDF galaxies\cite{AndXXV,Sextans_cluster,Eridanus_cluster,Caldwell} that may be clarified with deep velocity dispersion measurements. A wider $\psi$DM context may provide a natural
origin for such dense central stars clusters, and in general for the puzzling presence of nuclear star clusters commonly found in all types of galaxy, where a minority contribution to the universal DM from a heavier boson of $\simeq 10^{-20}$eV may sink within the wide solitons of the lighter, dominant DM (derived above of $10^{-22}$eV), resulting in a smaller dense DM structure that helps explain the puzzling origin and characteristic scale of nuclear star clusters \cite {Broadhurst2018}. This "multiple ultra-light" bosonic solution for
the Universal dark matter has the attraction of being underpinned theoretically by String Theory, where a wide discrete spectrum of axion-like particles extending to very light masses is generically predicted \cite{SW2006,Arvanitaki2010,Cicoli2012,vignozzi}, depending on the details of dimensional compactification.

\section*{Acknowledgements}
I.D.M. acknowledge the contribution of INFN (Iniziativa Specifica QGSKY).  This article is based upon work from COST Action CA1511 Cosmology and Astrophysics Network for Theoretical Advances and Training Actions (CANTATA), supported by COST (European Cooperation in Science and Technology)

\appendix

\section{\bf{$\psi$}DM Halo model}
Recently, numerical simulations of $\psi$DM model were carried out by Schive et al. (2014) \cite{Schive2014a}, 
and were subsequently used to show the presence of a cored dark matter region in the inner part of each virialized halo. This cored region of the dark matter halo satisfies the ground state of the Schroedinger-Poisson equations, and can be  described by \cite{Schive2014a}:
\begin{equation}\label{eq:sol_density}
\rho_c(r) \sim \frac{1.9~a^{-1}(m_\psi/10^{-23}~{\rm eV})^{-2}(r_c/{\rm kpc})^{-4}}{[1+9.1\times10^{-2}(r/r_c)^2]^8}~M_\odot {\rm pc}^{-3}.
\end{equation}
where $c_1=1.9$, $c_2=10^{-23}=$ , $c_3=9.1\times10^{-2}$ are constants;  and $r_c$ is the core radius which scales with the halo mass of the galaxy and the mass boson obeying  to the following the scaling relation which have been calibrated with cosmological simulations \cite{Schive2014b}:
\begin{equation}\label{eq:sol_radius}
r_c=1.6\biggl(\frac{10^{-22}}{m_b}\biggr)a^{1/2}
\biggl(\frac{\zeta(z)}{\zeta(0)}\biggr)^{-1/6}
\biggl(\frac{M_h}{10^9M_\odot}\biggr)^{-1/3}
\end{equation}

\section{Jeans Analysis}
In order to use the stellar kinematics of Antlia II to constrain the boson mass, $m_\psi$, 
we assume the galaxy to be spherically symmetric and supported by velocity dispersion. Then,  
we decompose the galaxy in the $\psi$DM halo and a stellar population which serves as tracer of 
the gravitational potential well dominated by the DM halo. Thus, the mass enclosed 
in a sphere of radius $r$ can be straightforwardly computed by integrating the density profile of eqn.~\ref{eq:sol_density} as
\begin{equation}\label{eq:massbulge1}
M_{DM}(r) = 4\pi\int_0^r x^2 \rho_{DM}(x)dx\,.
\end{equation}

Finally, the velocity dispersion can be obtained by integrating the Jeans equation 
\begin{equation}\label{eq:sol_Jeans}
\frac{d(\rho_*(r)\sigma_r^2(r))}{dr} = -\rho_*(r)\frac{M_{DM}(r)}{r^2}-2\beta\frac{\rho_*(r)\sigma_r^2(r)}{r},
\end{equation}
where $\beta \equiv 1 - \frac{\sigma_\phi^2 + \sigma_\theta^2}{2\sigma_r^2}$ is the anisotropy parameter, and
we have considered  stars with a Plummer density profile  
\begin{equation}
\rho_*(r) = \frac{3M_{*}}{4\pi r_{h}}  \biggl(1+\frac{r^2}{r_{h}^2}\biggr)^{-\frac{5}{2}}
\end{equation}

A general solution of eqn.~\ref{eq:sol_Jeans} for $\beta$ constant is given by [56] 
\begin{equation}
    \rho_*(r)\sigma_r^2(r)=Gr^{-2\beta}\int_r^\infty x^{2\beta-2}\rho_*(x)M_{DM}(x)dx\,,
\end{equation}
and, it must be projected along the line of sight to be compared with data:
\begin{equation}\label{eq:sol_projected}
\sigma^2_{los} (R) = \frac{2}{\Sigma (R)}\int_{R}^{\infty} \biggl(1-\beta \frac{R^2}{r^2}\biggr)\frac{\sigma^2_r(r)\rho_*(r)}{(r^2-R^2)^{1/2}} r dr\,.
\end{equation}

\section{Data, data analysis, and results}\label{app:C}

\begin{table*}[!ht]
\caption{In columns 2 and 3, are listed the half-light radius and the corresponding mass from \cite{walker2009,walker2010,Torrealba2018}. 
In column 4 and 5, we report the boson mass and soliton radius analyzed in \cite{Chen2017,Schive2014a}. 
In the columns 6 and 7 are the predicted values of the masses within the soliton radius and the virial radius, $M_{sol}$ and $M_{200}$, assuming a boson mass $~10^{-22}$ eV.} 
\centering
\begin{tabular}{lccccccc}
\hline
 {\bf Galaxy} & $r_h$ &$M(<r_h)$ &$m_\psi$ & $r_c$ & $M_{sol}$ & $M_{200}$  & {\bf Ref.}\\
              & (kpc) & ($10^7 M_\odot$)& ($10^{-22}$eV) & (kpc) & ($10^7 M_\odot$) & ($10^9 M_\odot$) & \\
 \hline
Antlia II & $2.90\pm0.31$& $5.4\pm2.1$ & [$0.6-1.4$] & [$1.8-3.4$] & $2.7$ & $0.19$ & This work \& \cite{Torrealba2018} \\
Fornax  & $0.67\pm0.34$& $5.3\pm0.9$& $0.81^{+0.16}_{-0.17}$ & $0.92^{+0.15}_{-0.11}$  &  $5.7$ & $1.1$ & \cite{walker2009,walker2010,Schive2014a}\\
Sextans  & $0.68\pm0.12$ & $2.5\pm0.9$& $2.13^{+1.08}_{-0.64}$ & $[1.5-3.0]$ & $4.6$ & $0.63$ & \cite{walker2009,walker2010,Chen2017}\\
Sculptor & $0.26\pm0.39$& $1.3\pm0.4$ & $1.23^{+0.41}_{-0.33}$ & $0.60^{+0.11}_{-0.12}$ & $6.9$ & $2.1$ & \cite{walker2009,walker2010,Chen2017}\\
Draco    & $0.20\pm0.12$ & $0.94\pm0.25$& $1.12\pm0.52$ & $0.56\pm0.13$ & $8.6$ & $4.3$ & \cite{walker2009,walker2010,Chen2017}\\
\hline
\end{tabular}
\end{table*}

The Antlia 2 dwarf galaxy has been recently discovered \cite{Torrealba2018}. 
It is located at the distance of 130 kpc from the Milky Way and extended $\sim 2.9 kpc$. This galaxy has the lowest measured surface brightness ($ 32.3 mag/arcsec^2$) that is $\sim100$ times more diffuse than the ultra diffuse galaxies.

 The satellite was recognised when combining photometric, astrometric
and variability information from the satellite {\em Gaia} in its second data release, while kinematic information were subsequently obtained 
with spectroscopic follow-up using AAOmega spectrograph on the Anglo-Australian Telescope which allowed a measurement of the velocity dispersion profile.

We have modeled these data within $\psi$DM picture described in the previous section. Retaining the observed values of the half light radius and the observed stellar mass within $r_h=2867\pm312$ pc \cite{Torrealba2018}, 
the theoretical counterpart is fully determined once the boson mass and the total halo mass are specified. Thus, we have explored the parameter space with the Monte Carlo Markov Chain (MCMC) pipeline employing the Metropolis-Hastings sampling  algorithm. We allowed for an adaptive step size in order to reach an acceptance rate between 20\% and 50\%, and we ensured the convergence relying on the Gelman-Rubin criteria.  We adopted flat priors covering the range $[0.1, 3]\times 10^{-22} eV$ for the boson mass, and $[0.5, 13]\times10^8 M_\odot$ for the total halo mass, while $\beta$ is set to zero [1], and we run four different chains with random starting points. Then, once the convergence criteria is satisfied,  the different chains are merged to compute the total likelihood, and  the 1D marginalized likelihood distribution with the corresponding the expectation value and  variance. 

\begin{figure}[H]
	\centering
	\includegraphics[width=0.98\columnwidth]{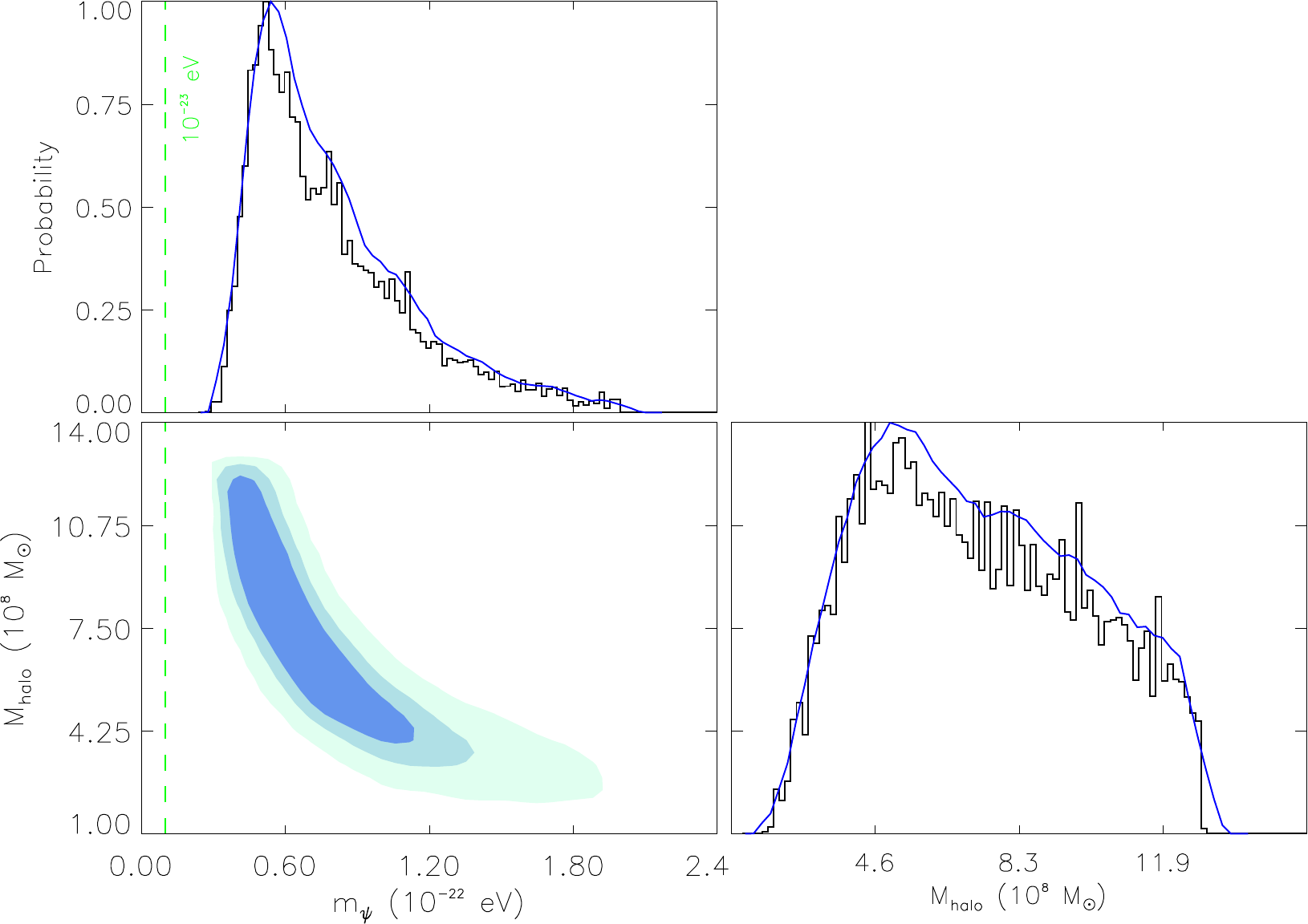}
	\caption{2D marginalized contours of the model parameters $[m_\psi, M_{halo}]$ obtained from our MCMC analysis of Antlia II.  The 68\%, 95\% and 99\% confidence levels are shown with the 1D marginalized likelihood distribution.}\label{fig4}
\end{figure}

The best fit values obtained with our analysis are $m_\psi (10^{-22} \rm{eV})=0.81^{+0.41}_{-0.21} $ and  $M_{halo}(10^8 M_\odot)=6.61^{+3.48}_{-2.68}$, respectively. The best fit value of the boson mass is consistent at $1\sigma$ with other analysis carried out in \cite{Schive2014a,Chen2017}. 
The 68\%, 95\%, and 99\% of confidence levels of the posterior distribution are shown in Fig. \ref{fig4}. We have also translated these constraints in a one on the core radius using Eq. \ref{eq:sol_radius}. We carried out 1000 Monte Carlo simulations randomly choosing the values of $m_\psi$ and $M_{halo}$ from their posterior distribution, and obtaining $r_c = 2.7\pm0.63$ kpc.

\begin{figure}[H]
	\centering
	\includegraphics[width=0.98\columnwidth]{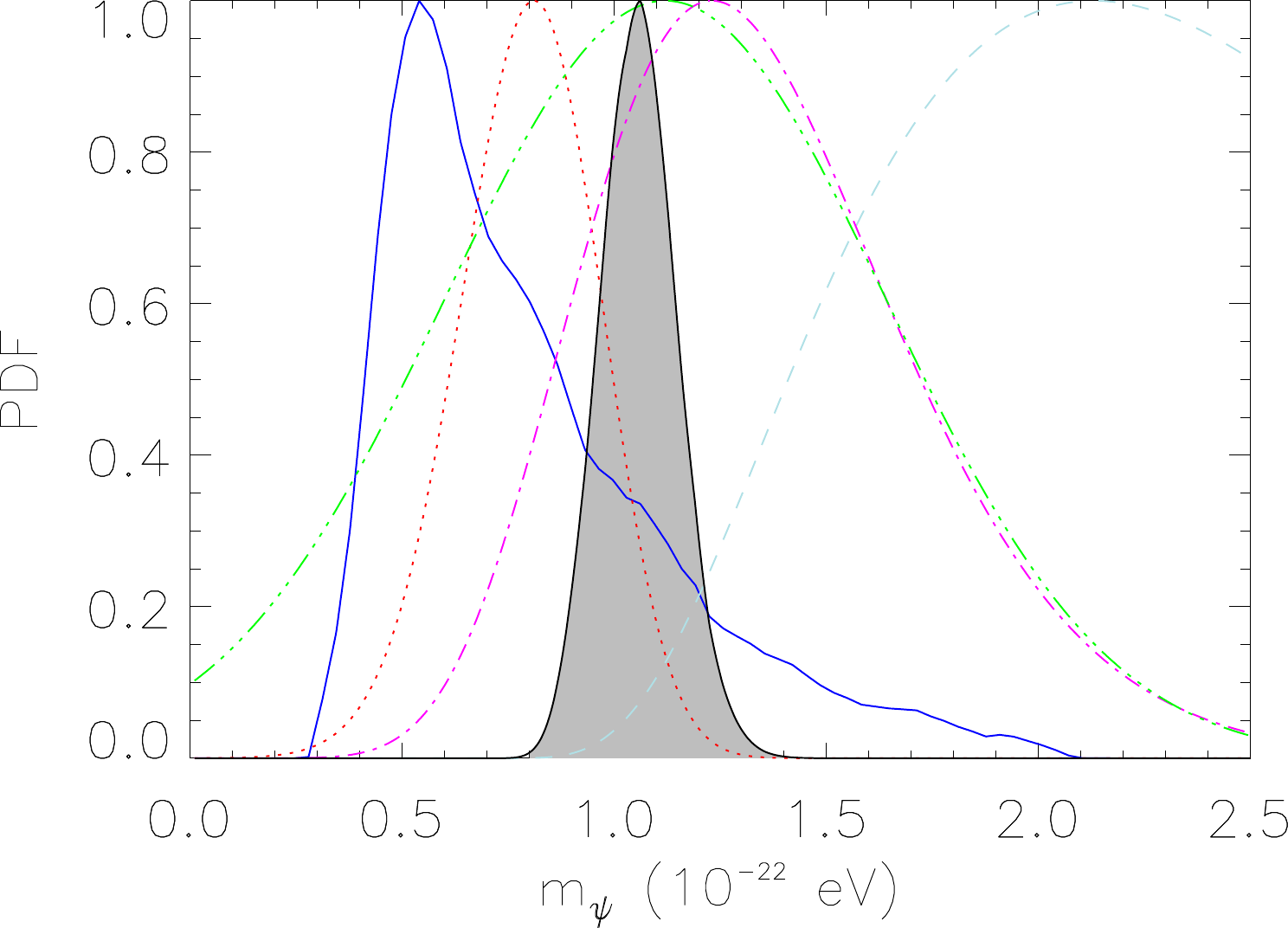}
	\caption{Joint likelihood constraint on the boson mass from our Antlia II analysis (solid blue line), together with previous published work on Fornax (dotted red), Sextans (dashed cyan), Sculptor (dot dashed magenta), Draco (triple dotted dashed green). The final joint probability distribution for the boson mass is shown as the black solid curve.}\label{fig5}
\end{figure}

Finally, in Fig. \ref{fig5}, we compare our constraint on the boson mass with other independent estimates (listed in Table 1) {assuming a gaussian posterior distribution}. From the comparison,  we obtain that a boson mass of $m_\psi=1.07\pm0.08$ for all the published boson mass analyses including our new measurement for Antlia II.

\section{Further Discussions and conclusions}
We have used the dispersion velocity of the stars in the newest discovered ghostly galaxy Antlia II to constrain the $\psi$DM model. Such a galaxy has pointed out several inconsistencies of the CDM model which may be solved by changing the DM paradigm to explain the existence of such a large stellar core in a barely visible galaxy. Our best estimation of the boson mass, namely $m_\psi (10^{-22} \rm{eV})=0.81^{+0.41}_{-0.21}$, is still comfortably consistent with other constraints coming from other independent analyses (see for example \cite{Chen2017,Lokas2003,Hlozek2018}).
Furthermore, as we use other constraints on $\psi$DM  coming from the analysis of the dispersion velocity profiles in other dwarf galaxies, we are able to estimate the boson mass $\sim8\%$ accuracy showing that the combination of galaxy data favor an ultralight boson with mass $\sim 10^{-22} \rm{eV}$. 

A previous analysis that used the UV-luminosity function to predict the reionization history of the universe found  a $3\sigma$ tension with boson masses $\sim 10^{-22}$ eV \cite{Bozek2015}.
Nevertheless, since they used the optical depth as measured by Planck  in the 2013 data release, those constraints are invalidated by the significant change of such a fundamental parameter in subsequent Planck results. Furthermore, from the existence of the star cluster in Eridanus II,  a boson mass of the order of $\sim 10^{-19}$ eV has been inferred \cite{2019PhRvL.123e1103M}.
Nevertheless, the existence of this star cluster may be explained by a multiple axion scenario as recently shown in \cite{Broadhurst2018}.

Finally, we note recent claims of significantly higher bounds on the boson mass obtained from the high frequency behaviour of the Ly-$\alpha$ power spectrum \cite{Baur2016}, 
made by analogy with Warm Dark Matter, under much debate  where differences in the matter power spectrum between $\psi$DM and WDM may play a critical role \cite{Hui2016, Zhang2018}. 
Dedicated simulations are really required for $\psi$DM that include hydrodynamics and cooling to  provide a more definite answer on whether or not Lyman-$\alpha$ excludes bosons with masses  $\sim 10^{-22}$ eV, as it is conceivable that excess variance in the small scale as distribution arises from the pervasive de- Broglie scale structure uncovered in the $\psi$DM simulations caused by the inherent interference in this context that fully modulates the density of halos and filaments, ranging from constructive to destructive interference. Furthermore, gas outflows and shocks
are observed to be widespread at high redshift that can also be expected to enhance the variance in power on small scales in the gas, but these are not included in the simulations relied on for Ly-$\alpha$ constraints even for Warm Dark Matter.

\end{document}